\documentclass[12pt,aps,prd,preprint,tightenlines,superscriptaddress,showpacs,nofootinbib]
{revtex4-1}
\usepackage{epsfig}
\usepackage{amssymb,amsmath}
\usepackage{hyperref}
\usepackage{xfrac}
\usepackage{subfig}   
\usepackage{color}
\usepackage{physics}
\usepackage[utf8]{inputenc} 
\usepackage{feynmp-auto}

\newcommand{\bea}{\begin{eqnarray}}
\newcommand{\eea}{\end{eqnarray}}
\begin{document}
%\begin{flushright}
%\end{flushright}
%
\vspace*{1.0cm}

\begin{center}
\baselineskip 20pt 
{\Large\bf 
Complementarity between gravitational wave signatures and Higgs precision measurements of a classically conformal hidden U(1) extended Standard Model
}
\vspace{1cm}

{\large
Victor Baules\footnote{vabaules@crimson.ua.edu} and Nobuchika Okada\footnote{okadan@ua.edu}
}
\vspace{.5cm}

{\baselineskip 20pt \it
Department of Physics and Astronomy, \\University of Alabama, Tuscaloosa, AL 35487, USA
} 

\vspace{.5cm}

\vspace{1.5cm} {\bf Abstract}
\end{center}

We consider a classically conformal extension of the Standard Model (SM) with a hidden $U(1)$ gauge symmetry, 
   where the $U(1)$ symmetry is radiatively broken via the Coleman-Weinberg mechanism. 
This radiative breaking then induces electroweak (EW) symmetry breaking through a negative mixed quartic coupling 
   between the hidden sector Higgs field and the SM Higgs doublet. 
Due to the mixed quartic coupling, the original two Higgs fields mix with a small angle $\theta$ 
   to form two mass eigenstates, $h_1$ (SM-like Higgs) and $h_2$ (SM singlet-like Higgs).  
Setting their masses as $M_{h_1} > 2 M_{h_2}$ to allow the decay process, $h_1 \to h_2 h_2$, 
   we find a remarkable prediction of the conformal Higgs potential: 
   the corresponding decay width is strongly suppressed in contrast to the one in the conventional Higgs potential. 
This anomalous behavior provides a striking experimental signature testable at the International Linear Collider (ILC) 
    and other future high-energy lepton colliders.
In this work, we investigate further experimental signatures of the model which are complementary to the Higgs physics at the colliders. 
First, we study the hidden $U(1)$ gauge boson as a cold dark matter (DM) candidate. 
While we find parameter regions consistent with both the observed relic abundance and collider sensitivities, 
   these regions are excluded by current direct DM detection limits. 
Second, we consider a strong first-order phase transition associated with hidden $U(1)$ breaking in the early universe 
   and compute the resulting gravitational wave (GW) spectrum. 
We find that for parameter regions complementary to Higgs precision measurements at the ILC, 
   the predicted GW signals lie well above the projected sensitivities of future GW observatories.

\thispagestyle{empty}

%\bigskip
\newpage

\addtocounter{page}{-1}

%%%%%%%%%%%%%%%%%%%%%%%%%%
%\baselineskip 36pt
% Main body
%%%%%%%%%%%%%%%%%%%%%%%%%%
%\baselineskip 18pt
%%%%%%%%%%%%%%%%%%%%%%%%%%

%%%%%%%%%%%%%%%%%%%%%%%%
\section{Introduction} 
\label{sec:1}
%%%%%%%%%%%%%%%%%%%%%%%%

Although the Standard Model (SM) of elementary particles has been highly successful,
   the origin of electroweak symmetry breaking (EWSB) remains an open question.
In the SM, a negative mass-squared term is introduced by hand into the Higgs potential 
   to trigger EWSB, leaving its fundamental origin unexplained. 

A well-known alternative is the Coleman-Weinberg (CW) mechanism \cite{Coleman:1973jx},
   where radiative corrections to the scalar potential from gauge interactions induce spontaneous symmetry breaking. 
However, this mechanism cannot be applied directly to the SM Higgs sector,
  since radiative corrections from the large top-quark Yukawa coupling render the potential unbounded from below \cite{Sher:1988mj}.
This limitation motivates extensions of the SM. 

The simplest extension introduces a new $U(1)$ gauge group, with classical conformal symmetry 
  enforcing the absence of mass terms at tree level (see, for example, Refs.~\cite{Iso:2009ss, Iso:2009nw}, and references thereof).
A new $U(1)$-charged Higgs field $\Phi$, singlet under the SM gauge groups, acquires a vacuum expectation value (VEV) 
   via the CW mechanism, thereby breaking the new gauge symmetry.
A negative mixed quartic coupling between $\Phi$ and the SM Higgs doublet $H$ then generates 
   an effective negative mass-squared term for $H$, driving EWSB.
In this picture, EWSB originates from the radiative breaking of the new $U(1)$ gauge symmetry.
This elegant structure motivates a detailed study of the Higgs phenomenology of the model.

In our previous work \cite{Baules:2022gva}, we investigated whether Higgs sectors with and without classical conformal symmetry 
  can be distinguished in SM extensions with a hidden $U(1)$ gauge group.
In the presence of the (small) mixed quartic coupling, $\Phi$ and $H$ mix with a (small) mixing angle $\theta$, 
  yielding two mass eigenstates: the SM-like Higgs boson $h_1$ and a new scalar $h_2$, mostly originating from $\Phi$. 
We focused on the case $M_{h_1} > 2 M_{h_2}$ for their mass eigenvalues, 
  where the Higgs anomalous decay $h_1 \to h_2 h_2$ is kinematically allowed.
Due to the mixing, the SM-like Higgs boson exhibits anomalous couplings.
For the same fixed value of the anomalous coupling in both models, we found a striking difference 
   in the behavior of the decay $h_1 \to h_2 h_2$. 
In the conventional Higgs potential (without classical conformal symmetry), the branching ratio of this decay scales 
   as $\sin^2 \theta$, as expected.
At the proposed International Linear Collider (ILC), both the anomalous Higgs coupling and the decay 
  can be observed for $\sin^2 \theta \geq 0.002$. 
By contrast, in the conformal Higgs potential, the branching ratio is strongly suppressed, scaling as $\sin^4 \theta$, 
  placing it well beyond the ILC sensitivity.
Therefore, if an anomalous Higgs coupling is measured at the ILC, the observation (or non-observation) of the decay 
   $h_1 \to h_2 h_2$ provides a sharp diagnostic: it would be expected in the conventional Higgs potential, 
   but absent if EWSB originates from the CW mechanism in a hidden $U(1)$ sector.

%%%%%%%%%%%%%%%%%%%%%%%%%%%%%%%%%%%%%%%%%%%%%%%%%%%%%%%%%%
% Add "The outline/plan of this paper is as follows" and roughtly outline what happens in which section.
%%%%%%%%%%%%%%%%%%%%%%%%%%%%%%%%%%%%%%%%%%%%%%%%%%%%%%%%%%

In this paper, we investigate additional experimental signatures of the classically conformal extension of the SM 
  with a hidden $U(1)$ gauge symmetry, which are complementary to Higgs boson precision measurements 
  at the ILC and other future high-energy lepton colliders.
We first examine the possibility that the hidden $U(1)$ gauge boson, denoted $Z'$, serves as a cold dark matter (DM) candidate. 
The $Z'$ boson can be stabilized by requiring its kinetic mixing with the SM $U(1)_{Y}$ gauge boson to vanish. 
We compute the relic abundance of $Z'$ DM through its dominant annihilation channel into a pair of $h_2$ bosons 
  and identify the parameter region consistent with the observed DM abundance. 
This parameter region partially overlaps with that identified in Ref.~\cite{Baules:2022gva} based on Higgs precision measurements at the ILC. However, in the region where the observed DM abundance and Higgs precision tests are complementary, 
  we find that the spin-independent $Z'$-nucleon scattering cross section, mediated by $h_{1,2}$ exchange, 
  exceeds current experimental bounds. 
Next, we consider a strong first-order phase transition (FOPT) associated with the breaking of the hidden $U(1)$ symmetry in the early universe, 
   and compute the resulting gravitational wave (GW) spectrum for the parameter space complementary to Higgs physics at the ILC. 
We find that for $30 \lesssim M_{h_2}[{\rm GeV}] \lesssim 50$ and $\sin^2 \theta = \mathcal{O}(0.01)$, 
   the predicted GW signal lies well above the projected sensitivities of future GW observatories, 
   while the same region remains accessible at the ILC.

This paper is organized as follows.
In Section \ref{sec:2}, we review the classically conformal $U(1)$ extension of the SM introduced in \cite{Baules:2022gva}, 
   and discuss future collider prospects for Higgs precision measurements.
In Section \ref{sec:3}, we investigate the possibility of the hidden $U(1)_H$ gauge boson serving as the sole DM candidate, 
   examining constraints from the observed relic abundance and direct detection experiments, 
   and highlighting their complementarity with future Higgs precision measurements.
In Section \ref{sec:4}, we explore an additional experimental probe: a possible strong first-order phase transition in the early universe, 
   which could generate a primordial GW background. 
We compute GW spectra for several experimentally allowed values of the scalar mass $m_{\phi}$ 
   and assess the detection prospects at both near- and far-future GW observatories.
Finally, Section \ref{sec:5} is devoted to our conclusions.

%%%%%%%%%%%%%%%%%%%%%%%%%%%%%%%%%%%%%%%%%%%%%%%%%%%%%%%%%%
\section{The Classically conformal $U(1)$ extended Standard Model}
\label{sec:2}
%%%%%%%%%%%%%%%%%%%%%%%%%%%%%%%%%%%%%%%%%%%%%%%%%%%%%%%%%%
% Make some modifications so that the wording/content isn't so similar to the previous paper, just in case - otherwise arXiv might flag for copying/plagarism! Just be sure!
This section is devoted to briefly review the key features of the classically conformal extended SM 
   and the Higgs boson phenomenology discussed in Ref.~\cite{Baules:2022gva}.
In this model the SM gauge group is extended to $SU(3)_{c} \times SU(2)_{L} \times U(1)_{Y} \times U(1)_{H}$, 
  where $U(1)_{H}$ is a newly introduced hidden sector.
The particle content of the model is extended only by the $U(1)_{H}$ gauge boson $Z^\prime$ and a new Higgs field  $\Phi$, 
  a singlet under the SM gauge group and with charge $Q_{\Phi}=+2$ under $U(1)_{H}$. 
Classically conformal invariance is imposed to forbid explicit mass terms in the Higgs potential, which at tree-level then reads
%%%%%%%%%%%%%%%%%%%%%%%%%%%%%%%%%%%%%%%%%%%%%%%%%%%%%%%%%%
\begin{equation}
\label{treelvV}
 V_{tree} = \lambda_{h}(H^{\dagger}H)^{2} + \lambda_{\phi} (\Phi^{\dagger}\Phi)^{2} - \lambda_{mix}(H^{\dagger}H)(\Phi^{\dagger}\Phi),
\end{equation}
%%%%%%%%%%%%%%%%%%%%%%%%%%%%%%%%%%%%%%%%%%%%%%%%%%%%%%%%%%
where $H$ is the SM Higgs doublet, and $\lambda_{mix}$ is taken to be positive. 
We choose to take $\lambda_{mix} \ll 1$, effectively separating the SM and $U(1)_{H}$ Higgs sectors.
At the tree-level, the vacuum of the scalar potential appears at the origin, and hence the EW and $U(1)_{H}$ symmetries remain unbroken.

In the $U(1)_{H}$ sector, we account for radiative corrections at the one-loop level \cite{Coleman:1973jx}
%%%%%%%%%%%%%%%%%%%%%%%%%%%%%%%%%%%%%%%%%%%%%%%%%%%%%%%%%%
\begin{equation}
V_{1-loop} = 
\frac{\beta_{\phi}}{8}  \left(\textnormal{ln}\left[\frac{\phi^2}{v_{\phi}^{2}}\right]-\frac{25}{6}\right) \phi^{4},
\end{equation}
%%%%%%%%%%%%%%%%%%%%%%%%%%%%%%%%%%%%%%%%%%%%%%%%%%%%%%%%%%
where $\phi = \sqrt{2} {\rm Re}[\Phi]$, and $\beta_{\phi}$ is given by
%%%%%%%%%%%%%%%%%%%%%%%%%%%%%%%%%%%%%%%%%%%%%%%%%%%%%%%%%%
\begin{equation}
\beta_{\phi} = \frac{1}{16 \pi^{2}}(20 \lambda_{\phi}^{2} + 96 g_{H}^{4}) \simeq \frac{1}{16 \pi^{2}}( 96 g_{H}^{4}),
\end{equation}
%%%%%%%%%%%%%%%%%%%%%%%%%%%%%%%%%%%%%%%%%%%%%%%%%%%%%%%%%%
where $g_{H}$ is the $U(1)_{H}$ gauge coupling, and we have assumed $\lambda_\phi^2 \ll g_H^4$. 
With these radiative corrections included, the $U(1)_{H}$ symmetry is spontaneously broken at $\langle \phi \rangle = v_{\phi}$. 
This is the CW mechanism.   
The full form of the potential is given by
%%%%%%%%%%%%%%%%%%%%%%%%%%%%%%%%%%%%%%%%%%%%%%%%%%%%%%%%%%
\bea
V = \frac{\lambda_{h}}{4} h^{4} + \left[ \frac{\lambda_{\phi}}{4} + \frac{\beta_{\phi}}{8}\left(\textnormal{ln}\left[\frac{\phi^2}{v_{\phi}^{2}}\right]-\frac{25}{6}\right) \right] \phi^{4}- \frac{\lambda_{mix}}{4} h^{2}\phi^{2}, 
\label{fullV}
\eea
%%%%%%%%%%%%%%%%%%%%%%%%%%%%%%%%%%%%%%%%%%%%%%%%%%%%%%%%%%
where $H = \frac{1}{\sqrt{2}}
\left(
\begin{matrix}
h \\
0
\end{matrix}
\right)$  
with $h$ a real scalar in the unitary gauge.
Note that the nonzero $\langle \phi \rangle$ generates a negative mass-squared term for $h$, and subsequently, 
  EWSB occurs at $\langle h \rangle = v_{h} = 246$ GeV. 
Therefore, one can identify the origin of EWSB as the radiative symmetry breaking in the hidden $U(1)_{H}$ sector 
  via the Coleman-Weinberg mechanism.

We analyze the full potential in Eq.~(\ref{fullV}) in the following way (see Ref.~\cite{Baules:2022gva} for more details): 
The potential includes five parameters, $\lambda_h$, $\lambda_\phi$, $\lambda_{mix}$, $\beta_\phi$ and $v_\phi$. 
Using the stationary conditions, $\frac{ \partial V }{ \partial \phi} \left|_{\phi = v_{\phi}, \, h=v_{h} }\right. = 0$ and  $\frac{ \partial V }{ \partial h} \left|_{\phi = v_{\phi}, \, h=v_{h} }\right. = 0$, we eliminate $\lambda_{\phi}$ and $\lambda_{mix}$ and express the potential in the following form:
\bea
V(h,\phi) &=&
\frac{1}{4} 
\left( 
	\frac{m_{h}^{2}}{2 v_{h}^{2}}
\right) h^{4} + \frac{1}{4} 
\left( \frac{11}{6}\beta_{\phi} + \frac{m_{h}^{2}v_{h}^{2}}{2v_{\phi}^{4}} \right)\phi^{4}  \nonumber \\
&& +  \frac{\beta_{\phi}}{8}
\left(
	 \textnormal{ln}\left[ 
		\frac{\phi^2}{v_{\phi}^{2}}
	\right] - \frac{25}{6} 
\right)\phi^{4} 
- \frac{1}{4}
\left( 
	\frac{m_{h}^{2}}{v_{\phi}^{2}}
 \right)h^{2}\phi^{2}, 
%\end{split}
\label{eq:VCW}
\eea
where $m_{h}^{2} \equiv 2 \lambda_{h} v_{h}^{2}$.
By shifting the fields $h\rightarrow h + v_{h}$ and $\phi \rightarrow \phi + v_{\phi}$, we can obtain the mass-squared matrix for the physical states,
\begin{equation}
 M_{sq} = 
\begin{pmatrix}
m_{h}^{2} & -M^{2} \\
- M^{2} & m_{\phi}^{2}
\end{pmatrix},
\label{mass-matrix}
\end{equation}
where $M^{2} = m_{h}^{2}\left( \frac{v_h}{v_{\phi}} \right)$ and $m_{\phi}^{2} \equiv  m_{h}^{2} \left( \frac{v_{h}}{v_{\phi}} \right)^{2} 
+ v_{\phi}^{2} \, \beta_{\phi}$.
The mass-squared matrix is diagonalized by
\begin{align}
\label{diag}
\left(
\begin{matrix}
h \\
\phi
\end{matrix}
\right)
=
\left(
\begin{matrix}
\cos(\theta) & \sin(\theta) \\
-\sin(\theta) &  \cos(\theta)   \\
\end{matrix}
\right)
\left(
\begin{matrix}
h_{1} \\
h_{2}
\end{matrix}
\right),
\end{align}
with a mixing angle $\theta$ defined by $\tan(2\theta) = \frac{2 M^{2}}{m_{h}^{2} - m_{\phi}^{2}}$, 
   where $h_{1}$ and $h_{2}$ are the mass eigenstates with mass eigenvalues $M_{h_{1}}$ and $M_{h_{2}}$, respectively.
For a small $\theta$ (or equivalently, a small $\lambda_{mix}$), 
   the mass eigenstate $h_1$ is the SM-like Higgs boson while $h_2$ is the SM singlet-like Higgs boson. 
In terms of observables ($M_{h_{1}}$, $M_{h_{2}}$, $\theta$), the parameters in Eq.~(\ref{mass-matrix})
can be expressed as follows:
\begin{eqnarray}
m_{h}^{2} &=& \frac{1}{2} \left(  M_{h_{1}}^{2} + M_{h_{2}}^{2} + \left(M_{h_{1}}^{2} - M_{h_{2}}^{2} \right) \cos(2 \theta) \right), \nonumber \\ 
m_{\phi}^{2} & \equiv &  m_{h}^{2} \left( \frac{v_{h}}{v_{\phi}} \right)^{2} + v_{\phi}^{2} \, \beta_{\phi}
= \frac{1}{2} \left( M_{h_{1}}^{2} +M_{h_{2}}^{2} - \left(M_{h_{1}}^{2} -M_{h_{2}}^{2} \right) \cos(2 \theta) \right), \nonumber \\
M^{2} &=& m_{h}^{2}\left( \frac{v_h}{v_{\phi}} \right) =
\frac{1}{2}  \left(M_{h_{1}}^{2} - M_{h_{2}}^{2} \right) \sin(2 \theta) . 
\label{relations}
\end{eqnarray}
Using these relations, the free parameters in the scalar potential of Eq.~(\ref{eq:VCW}), 
  namely $m_h$, $v_\phi$ and $\beta_\phi \simeq \frac{96}{16 \pi^2}g_H^4$, are expressed by $M_{h_{1}}$, $M_{h_{2}}$, and $\theta$. 
After setting $M_{h_1}=125$ GeV for the observed SM Higgs boson mass, 
  the Higgs potential is controlled by only two free parameters, $M_{h_2}$ and $\theta$. 
For $\theta \ll 1$, we can easily derive the following relations from Eq.~(\ref{relations}):
\bea
 && \theta \simeq \frac{v_h}{v_\phi} ,  \nonumber \\
 && g_H \simeq \left( \frac{\pi^2}{6} \beta_\phi \right)^{1/4} \simeq 
   \left( \frac{\pi^2}{6}\right)^{1/4} \left( \frac{M_{h_2}}{v_h} \right)^{1/2} \theta^{1/2} 
    \simeq 2  \left( \frac{\pi^2}{6}\right)^{1/2} \left( \frac{M_{h_2}}{m_{Z^\prime}} \right) ,  
\label{gHMzp}
\eea
where we have used the definition of $Z^\prime$ boson mass $m_{Z^\prime} =2 \, g_H \, v_\phi$ in the last expression. 
We are interested in the decay mode, $h_2 \to h_1 h_1$, caused by the trilinear coupling defined as 
   $g_{h_{1} h_{2} h_{2}} \equiv \frac{1}{2} \frac{\partial^{3} V}{\partial h_{1} \partial^{2} h_{2}} \left|_{h_{1} = h_{2} = 0}  \right.$. 
For $\theta \simeq \frac{v_h}{v_\phi} \ll1$, this coupling is approximately given by
%%%%%%%%%%%%%%%%%%%%%%%%%%%%%%%%%%%%%%%%%%%%%%%%%%%%%%%%%%
\bea
\label{CWTC}
g_{h_{1}h_{2}h_{2}}\Big|_{\rm conf} \simeq - \frac{M_{h_{2}}^{2}}{2  v_{h}}  \left(1 - 4 \frac{ M_{h_{2}}^{2}}{ M_{h_{1}}^{2}} \right)  \theta^{2}.
\eea
%%%%%%%%%%%%%%%%%%%%%%%%%%%%%%%%%%%%%%%%%%%%%%%%%%%%%%%%%%

Next, let us consider a similar extended Higgs potential without tree-level scale invariance for $h$ and $\phi$ in the unitary gauge:
\bea
V = \frac{1}{4} \hat{\lambda}_{h} \big( h^{2} - v_{h}^{2} \big)^{2} + \frac{1}{4} \hat{\lambda}_{\phi} \big(\phi^{2} - \hat{v}_{\phi}^{2} \big)^{2} + \frac{1}{4} \hat{\lambda}_{mix} \big( h^{2} - v_{h}^{2} \big) \big(\phi ^{2} - \hat{v}_{\phi}^{2} \big). 
\eea
We call this the ``conventional" potential, for simplicity, as this corresponds to a more conventional version of a Higgs-portal type potential.
It is obvious that the potential minimum appears at $\langle h \rangle = v_h$ and $\langle \phi \rangle = \hat{v}_\phi$. 
Using the stationary conditions, we rewrite the potential in terms of physical states 
   by shifting $h \rightarrow h + v_{h}$ and $\phi \rightarrow \phi + \hat{v}_{\phi}$:
\begin{align}
\label{eq:potnonconf}
V &= \frac{1}{4} \bigg( \frac{m_{h}^{2}}{2 v_{h}^{2}} \bigg) \big( h^{2} +2 h v_{h} \big)^{2} + \frac{1}{4} \bigg( \frac{m_{\phi}^{2}}{2 \hat{v}_{\phi}^{2}} \bigg) \big( \phi ^{2} + 2 \phi \hat{v}_{\phi} \big)^{2} + \frac{1}{4} \hat{\lambda}_{mix} \big( h ^{2} + 2 h v_{h} \big) \big( \phi ^{2} +2 \phi \hat{v}_{\phi} \big), 
\end{align}
where $m_h^2 =2 \lambda_h v_h^2$, and $m_\phi^2 =2 \hat{\lambda}_\phi \hat{v}_\phi^2$. 
We now read the mass-squared matrix for the conventional case of the form:
\begin{align}
M_{sq} &= 
\begin{pmatrix}
m_{h}^{2} &&M^{2}  \\
M^{2}        &&  m_{\phi}^{2}
\end{pmatrix},
\end{align}
where now $M^{2} = \hat{\lambda}_{mix} v_{h} \hat{v}_{\phi}$.
We diagonalize this matrix with Eq.~(\ref{diag}) with the mixing angle defined by $\tan(2 \theta) = - \frac{2 M^{2}}{m_{h}^{2} - m_{\phi}^{2}}$.
In terms of observables $M_{h_{1}}$, $M_{h_{2}}$, $\theta$, and $\hat{v}_{\phi}$, 
we can express the above mass matrix elements as 
\begin{align}
m_{h}^{2} &= \frac{1}{2} \left(  M_{h_{1}}^{2} + M_{h_{2}}^{2} + \left(M_{h_{1}}^{2} - M_{h_{2}}^{2} \right) \cos(2 \theta) \right), \nonumber  \\
m_{\phi}^{2} &= 2 \hat{\lambda}_\phi \hat{v}_\phi^2
= \frac{1}{2} \left( M_{h_{1}}^{2} +M_{h_{2}}^{2} - \left(M_{h_{1}}^{2} -M_{h_{2}}^{2} \right) \cos(2 \theta) \right), \nonumber \\
M^{2} &= \hat{\lambda}_{mix} v_{h} \hat{v}_{\phi} = -\frac{1}{2}\left(  M_{h_{1}}^{2} -  M_{h_{2}}^{2}\right) \sin(2 \theta).
\end{align}
Note that the conventional potential of Eq.~(\ref{eq:potnonconf}) is controlled by four free parameters in contrast to our conformal model. 
After setting $M_{h_1}=125$ GeV, the potential is controlled by three parameters: $M_{h_2}$, $\theta$ and $\hat{v}_{\phi}$.

The trilinear coupling $g_{h_{1} h_{2} h_{2}}$ for the conventional case is given by
\begin{align}
g_{h_{1} h_{2} h_{2}} \Big|_{\rm conv}= \frac{ \left( M_{h_{1}}^{2} + 2 M_{h_{2}}^{2} \right) \left( - v_{h} \cos(\theta) + \hat{v}_{\phi} \sin(\theta) \right) \sin(2 \theta)}{ 2 v_{h} \hat{v}_{\phi} }. 
\end{align}
This coupling has different behavior for two regions, $\theta \ll \frac{v_{h}}{\hat{v}_{\phi}}  \ll 1$ and $\frac{v_{h}}{\hat{v}_{\phi}}  \lesssim \theta  \ll 1$.
For $\theta \ll \frac{v_{h}}{\hat{v}_{\phi}}$, the trilinear coupling $g_{h_{1} h_{2} h_{2}}$ is approximately 
\begin{equation}
\label{nonCWTCsmall}
g_{h_{1} h_{2} h_{2}} \Big|_{\rm conv} \simeq - \frac{M_{h_{1}}^{2}}{2 \hat{v}_{\phi}} \left(1 + 2 \frac{M_{h_{2}}^{2}}{M_{h_{1}}^{2}} \right) \theta, 
\end{equation}
while for $ \frac{v_{h}}{\hat{v}_{\phi}} \lesssim \theta $, the coupling is 
\begin{equation}
\label{nonCWTCbig}
g_{h_{1} h_{2} h_{2}} \Big|_{\rm conv} \simeq  \frac{M_{h_{1}}^{2}}{2 v_{h}} \left(1 + 2 \frac{M_{h_{2}}^{2}}{M_{h_{1}}^{2}} \right) \theta^{2}. 
\end{equation}
As discussed, $\theta \simeq \frac{v_h}{v_\phi}$ in the conformal model for $\theta \ll1$. 
Although $\hat{v}_{\phi}$ is independent of $\theta$ in the conventional model, we may consider the case of $\hat{v}_{\phi} = v_\phi$ 
   to see a difference between the trilinear couplings in the two models. 
Since the decay width of the process $h_1 \to h_2 h_2$ is proportional to the square of the trilinear coupling, 
   Eqs.~(\ref{CWTC}) and (\ref{nonCWTCbig}) leads to the following ratio of the decay widths: 
\begin{equation}
\frac{\Gamma(h_1 \to h_2 h_2)|_{\rm conf}}{\Gamma(h_1 \to h_2 h_2)|_{\rm conv} } 
=\left( \frac{g_{h_{1} h_{2} h_{2}\Big|_{\rm conf}}}{g_{h_{1} h_{2} h_{2}\Big|_{\rm conv}}} \right)^2 
\simeq  \left(\frac{ M_{h_{2}}^{2}}{ M_{h_{1}}^{2} } \right)^2 \left( \frac{  1 - 4 \frac{ M_{h_{2}}^{2}}{ M_{h_{1}}^{2}}}{1 + 2 \frac{M_{h_{2}}^{2}}{M_{h_{1}}^{2}}  } \right)^2 <  2.6 \times 10^{-2}
\label{ratio}
\end{equation}
for $M_{h_{2}} < \frac{1}{2} M_{h_{1}}$, with the maximum value obtained for $M_{h_{2}} \simeq 0.33 \, M_{h_{1}}$. 
Therefore, for the same values for $M_{h_2}$ and $\theta$, the partial decay width in the conformal model is highly suppressed 
   compared to the conventional model. 
Note that in both models, the Higgs anomalous coupling is proportional to $\cos(\theta)$. 
The big difference between the partial decay widths has interesting implications for Higgs phenomenology at the ILC, 
   namely, even if the anomalous SM Higgs coupling is detectable in size,
   the SM-like Higgs boson $h_1$ decay to $h_2 h_2$ can be much harder to detect in our model. 
This suppression is a striking property of our conformal model, in which EWSB is triggered by the radiative $U(1)_{H}$ symmetry breaking.

Let us now consider outcomes for experimental probes of our model.
The Higgs physics of interest here are the anomalous Higgs decay mode $h_{1} \rightarrow h_{2} h_{2}$ and 
   the Higgs anomalous coupling proportional to $\cos(\theta)$. 
The partial decay width for the process $h_{1} \rightarrow h_{2} h_{2}$ is given by 
\bea
\Gamma(h_{1} \rightarrow h_{2} h_{2}) = \frac{|g_{h_{1} h_{2} h_{2}}|^{2}}{8 \pi M_{h_{1}}} \sqrt{ 1 - 4 \frac{M_{h_{2}}^{2}}{M_{h_{1}}^{2}} }.
\eea
Using the SM-like Higgs boson total decay width to SM particles $\Gamma_{SM} \simeq  4.1$ MeV \cite{Dittmaier:2012vm} 
  for $M_{h_1}=125$ GeV, the branching ratio to a pair of $h_{2}$ is given by
\bea
{\rm Br} \left(h_{1} \rightarrow h_{2} h_{2} \right) = \frac{\Gamma(h_{1} \rightarrow h_{2} h_{2})}{\Gamma_{SM} \cos^2\theta +\Gamma(h_{1} \rightarrow h_{2} h_{2})} \simeq  \frac{\Gamma(h_{1} \rightarrow h_{2} h_{2})}{\Gamma_{SM}} 
\eea
for $\theta \ll 1$.

%%%%%%%%%%%
\begin{figure}[h!]
\includegraphics[scale=1.0]{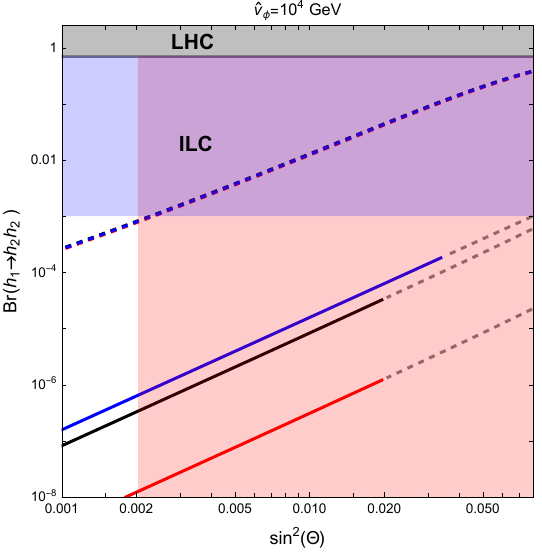}
\caption{
Comparison of conventional (colored dashed lines) and conformal (colored solid lines) ${\rm Br}(h_{1} \rightarrow h_{2} h_{2})$ 
   for $M_{h_{2}} = 10$ (red), 25 (black), and 50 (blue) GeV. 
Shaded regions correspond to the search reach of the proposed ILC: blue for the Higgs anomalous branching ratio \cite{Liu:2016zki, Yamamoto:2021kig, Fujii:2015jha}, and red for the Higgs anomalous coupling in terms of $\sin^2 (\theta)$ \cite{Barklow:2017suo, Bambade:2019fyw}.
The grey shaded regions are excluded by ATLAS($h_{1} \rightarrow h_{2} h_{2}\rightarrow b \bar{b} b \bar{b}$) \cite{ATLAS:2020ahi, ATLAS:2024itc} and the grey dashed lines by LEP-II \cite{Sirunyan:2018koj}.}%, Barate:2003sz}.}
% UPDATE CORRECT LEP-II REFERENCE HERE, NOT CMS.
\label{fig:BrOfSinsq}
\end{figure}
%%%%%%%%%%%%%%

In Fig.~\ref{fig:BrOfSinsq}, we present the branching ratio ${\rm Br}(h_{1} \rightarrow h_{2} h_{2})$ 
     in the conformal model for $M_{h_{2}} = 10$ (red), 25 (black), and 50 (blue) GeV (solid lines). 
The dashed lines show the corresponding results in the conventional model, where the results for three mass choices overlap indistinguishably.
Here, we have set $\hat{v}_{\phi}=10$ TeV for the conventional model. 
As expected, ${\rm Br}(h_{1} \rightarrow h_{2} h_{2})$ is highly suppressed for the same choice of $\sin^2 (\theta)$. 
The grey shaded region is excluded from the search for a Higgs anomalous decay $h_{1} \rightarrow h_{2} h_{2} \rightarrow b \bar{b} b \bar{b}$ 
   by the ATLAS collaboration \cite{ATLAS:2020ahi, ATLAS:2024itc}.
The SM singlet-like Higgs boson with our choice of mass mainly decays to $b \bar{b}$ through its mixing with the SM Higgs doublet.
The regions depicted by the grey dashed lines for  $M_{h_{2}} = 10, 25, \textrm{ and } 50$ GeV are excluded by the LEP-II results 
   for the Higgs boson search \cite{Barate:2003sz}. 
The prospective ILC search reaches are represented by the blue shaded region corresponding to the branching ratio\cite{Liu:2016zki, Yamamoto:2021kig, Fujii:2015jha}, and red shaded region for the Higgs anomalous coupling \cite{Barklow:2017suo, Bambade:2019fyw}.
Likewise, a muon collider (MuC) would be able to probe for the same signals as the ILC, 
   but with higher sensitivity and potential search reach. 
% As of yet, no extensive formal studies have been done for Higgs anomalous decays, but some predictions for MuC sensitivity can give a very general idea of MuC prospects.
For a $10 \, \textrm{TeV}$ MuC, the expected precision of Higgs width measurements is around $\mathcal{O} \left( 0.1 \% \right)$ \cite{InternationalMuonCollider:2025sys}, giving roughly the same anticipated search reach as the ILC down to roughly ${\rm Br}(h_{1} \rightarrow h_{2} h_{2}) \sim 10^{-3} $ for the $b \bar{b} b \bar{b}$ final state \cite{Fan:2026yuh}.
%A more optimistic claim is that a $30 \, \textrm{TeV}$ MuC may be sensitive to exotic Higgs decays with branching fractions as low as $\mathcal{O} \left( 10^{-7} \right)$. \cite{AlAli:2021let, Franceschini:2021aqd}.
For the anomalous coupling, the claimed precision is similar to the branching ratio sensitivity, down to around $\mathcal{O} \left( 0.1 \% \right)$ \cite{Forslund:2023reu}.

The combination of anomalous Higgs decay and anomalous Higgs coupling measurements provides a means 
   to distinguish our conformal scenario from the conventional model. 
In the conventional case, both the anomalous decay branching ratio and the anomalous coupling can fall within the ILC sensitivity
   for $\sin^2\theta \gtrsim 0.02$, allowing them to be measured simultaneously.  
In contrast, in our conformal model the branching ratio is strongly suppressed; thus, even if the anomalous coupling is observed, 
  the anomalous decay mode will still evade detection.
This feature offers an important clue to the origin of electroweak symmetry breaking (EWSB).

%%%%%%%%%%%%%%%%%%%%%%%%%%%%%%%%%%%%%%%%%%%%%%%%%%%%%%%%
\section{Hidden $Z'$ Vector Boson Dark Matter}
\label{sec:3}
%%%%%%%%%%%%%%%%%%%%%%%%%%%%%%%%%%%%%%%%%%%%%%%%%%%%%%%%
As a simple example of the BSM physics potential of our conformal model, we consider the possibility of 
  the $U(1)_H$ gauge boson $Z^\prime$ as the DM candidate. 
To ensure the stability of $Z^\prime$ boson, we assume no kinetic mixing between the $U(1)_{H}$ and the SM $U(1)_{Y}$ gauge bosons. 
Through its couplings with $h_{1,2}$, the $Z^\prime$ boson DM is in thermal equilibrium in the early universe, and thus 
   the DM relic abundance at present day is set via the freeze-out mechanism \cite{Kolb:1990vq}.  
In this scenario, the DM relic density is obtained by solving the Boltzmann equation: 
%%%%%%%%%%%%%%%%%%%%%%%%%%%%%%%%%%%%%%%%%%%%%%%%%%%%%%%%%%
\begin{equation}
    \frac{dY}{dx} = -\frac{s\left( T=m \right)}{H \left( T=m \right)} \frac{\langle \sigma v_{rel} \rangle}{x^{2}} \left( Y^{2} - Y_{EQ}^{2} \right),
\end{equation}
%%%%%%%%%%%%%%%%%%%%%%%%%%%%%%%%%%%%%%%%%%%%%%%%%%%%%%%%%%
where $x = m/T$ is the ratio between the DM mass $m$ and temperature $T$, 
   $Y = n/s$ is the yield defined by the ratio of the DM number density $n\left( T \right)$ to the entropy density of the universe 
    $s\left( T \right) = \left( 2\pi^{2}/45 \right) g_{*} T^{3}$ with $g_{*} \simeq 100$ being the effective total number of relativistic degrees of freedom, 
    $Y_{EQ}$ being the yield at thermal equilibrium, Hubble parameter $H\left( T \right) = \sqrt{\pi^{2} g_{*}/90} \left( T^{2}/M_{P} \right)$ 
    with reduced Planck mass $M_{P} = 2.43 \times 10^{18}$ GeV, and $\langle \sigma v_{rel} \rangle$ the thermal-averaged DM pair annihilation   
    cross section times DM relative velocity. 
With a chosen DM model, $\langle \sigma v_{rel} \rangle$ can be calculated as a function of $x$ and the present DM yield 
    $Y\left( x \rightarrow \infty \right)$ evaluated by solving the Boltzmann equation 
    using the initial condition $Y = Y_{EQ}$ for $x \ll 1$.
The relic density of DM is then given by 
%%%%%%%%%%%%%%%%%%%%%%%%%%%%%%%%%%%%%%%%%%%%%%%%%%%%%%%%%%
\begin{equation}
    \Omega_{DM} h^{2} = \frac{m Y\left( \infty \right) s_{0}}{\rho_{c} / h^{2}}    
\end{equation}
%%%%%%%%%%%%%%%%%%%%%%%%%%%%%%%%%%%%%%%%%%%%%%%%%%%%%%%%%%
where $s_{0} = 2890$ cm$^{-3}$ is the entropy density of the universe at present, 
   and $\rho_{c}/h^{2} = 1.05 \times 10^{-5} \,  \text{GeV/cm}^{3}$ is the critical density. 
We assume the relic $Z^\prime$ boson density constitutes the entire DM density observed at present, 
    $\Omega_{DM} h^{2}  =0.12$ \cite{Planck:2018vyg}.

\begin{figure}[ht!]
    \centering
    \includegraphics{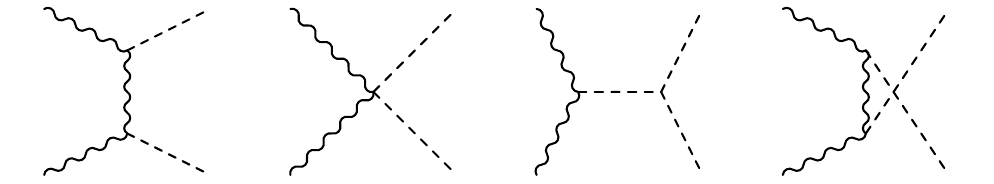}
    \caption{The four $Z'Z'\rightarrow h_{2} h_{2}$ annihilation processes.}
    \label{fig:DM_ann_processes}
\end{figure}

There are four processes of interest through which the dark matter annihilation $Z^\prime Z^\prime \rightarrow h_{2} h_{2}$ proceeds, 
  as illustrated in Fig.~\ref{fig:DM_ann_processes}. 
Since the mixing between $\Phi$ and $H$ is very small, $Z^\prime$ pair annihilation processes involving $h_1$ are negligibly small 
  and are therefore ignored in our calculation. 
The annihilation processes are dominated by $s$-wave, so the thermal-averaged cross section times relative velocity 
  can be well approximated by the zero-temperature cross section:
%%%%%%%%%%%%%%%%%%%%%%%%%%%%%%%%%%%%%%%%%%%%%%%%%%%%%%%%%%
\bea
\label{thermalsv}
\langle \sigma v_{\rm rel} \rangle ,[{\rm pb}] \simeq 9.75 \times 10^{8} \, \frac{g_{H}^{4}}{m_{Z^\prime}^{2}} , 
\eea
%%%%%%%%%%%%%%%%%%%%%%%%%%%%%%%%%%%%%%%%%%%%%%%%%%%%%%%%%%
where we have neglected $M_{h_2}$ and $[m_{Z'}] = \text{GeV}$. 
It is well known that the observed relic density, $\Omega_{\rm DM} h^{2} \simeq 0.12$, 
  is obtained when $\langle \sigma v_{\rm rel} \rangle \simeq 1$ pb. 
Imposing $\langle \sigma v_{\rm rel} \rangle [{\rm pb}] = 1$,  Eq.~(\ref{thermalsv}) leads to a relation between $g_H$ and $m_{Z^\prime}$:
\bea  
  g_{H} \simeq 5.2 \times 10^{-3} \sqrt{\frac{m_{Z^\prime}}{\rm GeV}} .
\label{DMdensity}
\eea

\begin{figure}[t!]
\begin{center}
%[width = 0.47\textwidth]
\includegraphics[width=0.40\textwidth]{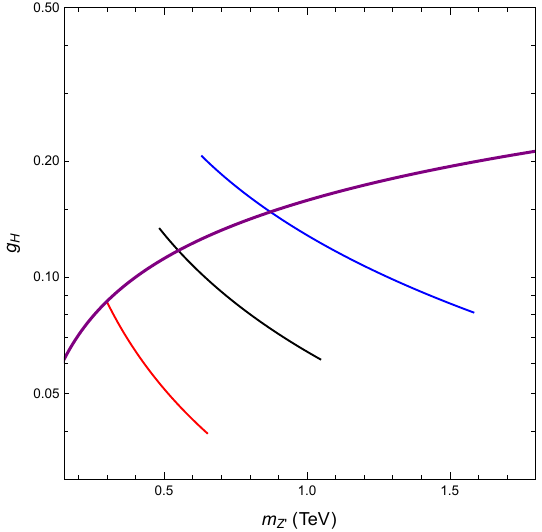} \; \; 
\includegraphics[width=0.45\textwidth]{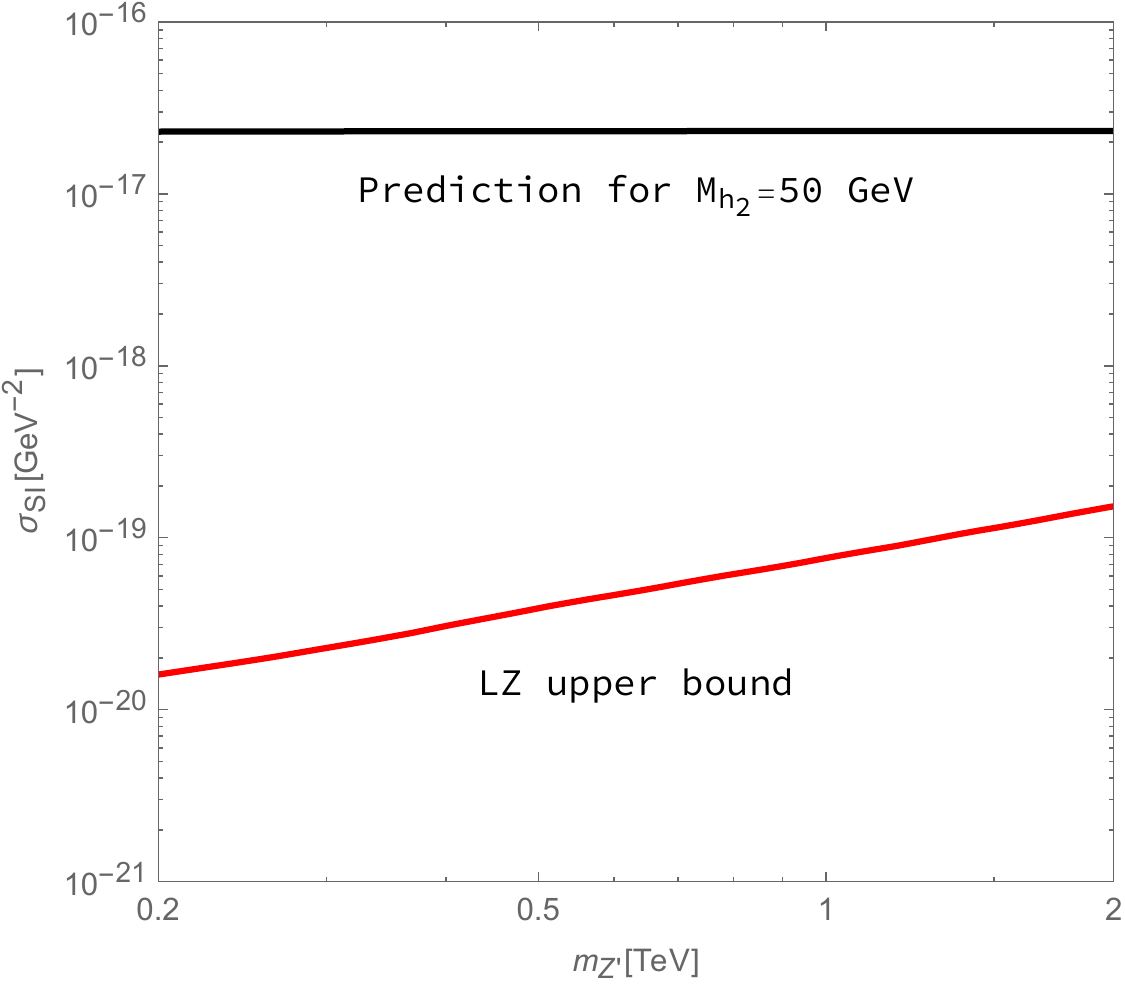}
\caption{
{\it Left Panel}: 
Complementarity of Higgs phenomenology at the ILC and the $Z^\prime$ DM relic density constraint in the $(m_{Z^{\prime}}$, $g_{H})$-plane. 
The purple line indicates the parameter sets that yield the observed DM relic abundance. 
The three diagonal solid lines in red, black, and blue (from left to right) correspond to the benchmark scenarios 
   with the same color coding as in Fig.\ref{fig:BrOfSinsq}. 
Each line is plotted over the range of $\sin^2\theta$ between the LEP-II upper bound (top end point) 
   and the prospective ILC search reach ($\sin^2\theta = 0.002$). 
{\it Right Panel}: Comparison of DM Direct detection cross section for $M_{h_{2}} = 50$ GeV 
and LZ direct detection bounds. Severity of the LZ bounds excludes vector DM in our conformal scenario.
}
\label{fig:Zpparamspace}    
\end{center}
\end{figure}

In the left panel of Fig.\ref{fig:Zpparamspace}, we examine the interplay between benchmark predictions of the conformal model 
   for Higgs phenomenology at the ILC (as discussed in the previous section) and the requirement of reproducing the observed DM relic density, 
   shown in the $\left(m_{Z'}, g_{H}\right)$ plane. 
The purple line indicates the parameter sets that yield the observed DM relic abundance. 
The three diagonal solid lines in red, black, and blue (from left to right) correspond to the benchmark scenarios 
   with the same color coding as in Fig.\ref{fig:BrOfSinsq}.
As discussed earlier, once the scalar mass $M_{h_2}$ is fixed, the coupling $g_H$ is determined as a function of $m_{Z'}$ 
  in our conformal model (see Eq.~\ref{gHMzp} for an approximate analytic expression). 
Each line is plotted over the range of $\sin^2\theta$ between the LEP-II upper bound (top end point) 
   and the prospective ILC search reach ($\sin^2\theta = 0.002$). 
The intersection points between the colored benchmark lines and the purple line identify the parameter sets 
  for which Higgs phenomenology at the ILC and DM physics serve as complementary probes of the model. 
It is also noteworthy that for $M_{h_2} < 10$ GeV, no such intersections occur, 
  so that the two sets of constraints cannot be simultaneously satisfied in this mass range.

%%%%%%%%%%%%  Fig  %%%%%%%%%%%%%%%%%%%%%%%%%%%%%%%%%%%%%%
%%%%%%%%%%%%%%%%%%%%%%%%%%%%%%%%%%%%%%%%%%%%%%%%%%%%%

Although the mixing $\Phi$ and $H$ is very small, the DM $Z^\prime$ boson couples to the SM fermions through this mixing. 
The direct DM detection experiments have been searching for the signal of the elastic DM matter scattering off nuclei, 
   and the null results provides a very severe upper bound on the spin-independent cross section of a DM particle 
   with a nucleon $\sigma_{SI}$.  
Here, we calculate $\sigma_{SI}$ for the DM $Z^\prime$ boson mediated by $h_{1,2}$. 

The coupling of $Z^\prime$ with $h_{1,2}$ is extracted from the kinetic term of $\Phi$ in the Lagrangian: 
%%%%%%%%%%%%%%%%%%%%%%%%%%%%%%%%%%%%%%%%%%%%%%%
%\begin{equation}
%    \mathcal{L} \supset - \frac{1}{4} Z^{'}_{\mu \nu} Z^{' \mu \nu} + \left( D_{\mu} \Phi \right)^{\dagger} \left( D^{\mu} \Phi \right)
%\end{equation}
\bea
 \mathcal{L} \supset  \left( D_{\mu} \Phi \right)^{\dagger} \left( D^{\mu} \Phi \right), 
\eea
%%%%%%%%%%%%%%%%%%%%%%%%%%%%%%%%%%%%%%%%%%%%%%%
where $D_{\mu} = \partial_{\mu} - i2g_{H} Z^\prime_{\mu}$ is the covariant derivative. 
%and $Z^{'}_{\mu \nu} \equiv \partial_{\mu} Z^{'}_{\nu} + \partial_{\nu} Z^{'}_{\mu}$.
Substituting $\Phi = \frac{1}{\sqrt{2}} \left( v_{\phi} + \phi \right)$, we have
%%%%%%%%%%%%%%%%%%%%%%%%%%%%%%%%%%%%%%%%%%%%%%%
\bea
    \mathcal{L} &\supset  \left( D_{\mu} \Phi \right)^{\dagger} \left( D^{\mu} \Phi \right) 
    \supset 4 \, g_{H}^{2} \, v_{\phi} \, \phi \, Z^\prime_{\mu} Z^{\prime \mu}. 
\eea
%%%%%%%%%%%%%%%%%%%%%%%%%%%%%%%%%%%%%%%%%%%%%%%
In the original basis, the DM $Z^\prime$ boson has no direct interaction with the SM particles, except for the SM Higgs, 
   while the SM Higgs boson ($h$) has couplings with the SM quarks and an effective couplings with a pair of gluons. 
One then finds that terms relevant to direct DM detection interactions are:
%%%%%%%%%%%%%%%%%%%%%%%%%%%%%%%%%%%%%%%%%%%%%%%
\begin{equation}
    \mathcal{L}_{int} \supset 4 g_{H}^{2} v_{\phi} \phi Z^\prime_{\mu} Z^{\prime \mu}
     - \left( \sum_{q} \frac{m_{q}}{v_{h}}\bar{q} q - \frac{\alpha_{s}}{4 \pi v_{h}} G_{\mu \nu} G^{\mu \nu} \right) h. 
\end{equation}
%%%%%%%%%%%%%%%%%%%%%%%%%%%%%%%%%%%%%%%%%%%%%%%
We now express these interactions by the mass eigenstates $h_{1,2}$:
\bea
    \mathcal{L}_{int} &\supset & 
    h_{1} \left\{ - 4 g_{H}^{2} v_{h} s_{\theta} Z^\prime_{\mu} Z^{\prime\mu} - \left( \sum_{q} \frac{m_{q}}{v_{h}}\bar{q} q - \frac{\alpha_{s}}{4 \pi g_{H}} G_{\mu \nu} G^{\mu \nu}  \right) \cos{\theta} \right\} \nonumber \\
 & & + h_{2} \left\{ 4 g_{H}^{2} v_{h} c_{\theta} Z^\prime_{\mu} Z^{\prime \mu} - \left( \sum_{q} \frac{m_{q}}{v_{h}}\bar{q} q - \frac{\alpha_{s}}{4 \pi v_{h}} G_{\mu \nu} G^{\mu \nu}  \right) \sin{\theta} \right\} 
%    \mathcal{L}_{mass} &= - \frac{1}{2} M_{h_{1}} h_{1}^{2} - \frac{1}{2} M_{h_{2}} h_{2}^{2}.
\eea

For the elastic scattering process $Z^\prime q \rightarrow Z^\prime q$, which is relevant for direct DM detection, 
   the momentum transfer is much smaller than $M_{h_{1}}, M_{h_{2}}$.  
After integrating out $h_{1,2}$, we obtain an effective Lagrangian:
\bea
    \mathcal{L}_{int} &=& \frac{c_{1}}{M^{2}_{h_{1}}} Z^\prime_{\mu}Z^{\prime \mu} \left( \sum_{q} \frac{m_{q}}{v_{h}}\bar{q} q - \frac{\alpha_{s}}{4 \pi v_{h}} G_{\mu \nu} G^{\mu \nu} \right) \nonumber \\
   &&+ \frac{c_{2}}{M^{2}_{h_{2}}} Z^\prime_{\mu}Z^{\prime \mu} \left( \sum_{q} \frac{m_{q}}{v_{H}}\bar{q} q - \frac{\alpha_{s}}{4 \pi v_{h}} G_{\mu \nu} G^{\mu \nu} \right), 
\eea
where $c_{1,2} = \pm 8 g_{H} \left( v_{\phi} / v_{h} \right) \sin{\theta} \cos{\theta} \simeq \pm 8 g_{H}$ for $\theta \simeq v_{h}/v_{\phi} \ll 1$.
Using the effective Lagrangian above, the scattering cross section between the DM $Z^\prime$ boson and a nucleon $N$ 
  can be given by (see, for example, Ref.~\cite{Kanemura:2010sh}):
\bea
    \sigma_{SI} \left( Z^\prime N \rightarrow Z^\prime N \right) 
    % \left( \frac{c_{1}}{2 M_{h_{1}^{2}} + c_{2}}{2 M_{h_{2}}^{2}} \right)^{2} \left( \frac{m_{N}}{m_{Z^{'}} + m_{N}} \right)^{2} \frac{f_{N}^{2}}{\pi}\\
 =\frac{16 g_{H}^{4}}{\pi} \left( \frac{1}{M_{h_{1}}^{2}} - \frac{1}{M_{h_{2}}^{2}} \right)^{2} \left( \frac{m_{N}}{m_{Z^\prime} + m_{N}} \right)^{2}f_{N}^{2}, 
\eea
where
\begin{equation}
    f_{N} = \sum_{q} m_{q} \bra{N} \bar{q} q \ket{N} - \frac{\alpha_{s}}{4 \pi} \bra{N} G_{\mu \nu} G^{\mu \nu} \ket{N} \simeq 0.25 \, \text{GeV}.
\end{equation}

To reproduce the observed DM relic abundance, $g_H$ and $m_{Z^\prime}$ must satisfy the relation in Eq.~(\ref{DMdensity}). 
This implies the spin-independent cross section, $\sigma_{SI} \left( Z^\prime N \rightarrow Z^\prime N \right)$, 
  becomes nearly independent of the DM mass $m_{Z^\prime}$ for $m_{Z^\prime} \gg m_{N} = 0.938$ GeV.  
In the right panel of Fig.~\ref{fig:Zpparamspace}, we show $\sigma_{SI}$ for $M_{h_{2}} = 50$ GeV 
  alongside the latest LUX-ZEPLIN (LZ) upper bound \cite{LZ:2024zvo}. 
Unfortunately, the $Z^\prime$ DM scenario is excluded by the LZ results. 
On the other hand, if the $Z^\prime$ boson has a nonzero kinetic mixing with the SM hypercharge gauge boson, 
  it becomes unstable, rendering its existence cosmologically harmless.

%%%%%%%%%%%%%%%%%%%%%%%%%%%%%%%%%%%%%%%%%%%%%%%%%%%%%%%%
\section{Gravitational Wave signals from conformal first order phase transitions}
\label{sec:4}
%%%%%%%%%%%%%%%%%%%%%%%%%%%%%%%%%%%%%%%%%%%%%%%%%%%%%%%%
As a further probe of complementarity, we investigate the possibility of GW signals 
   arising from a FOPT associated with the breaking of the $U(1)_{H}$ gauge symmetry in the early universe. 
In this section, we first provide a brief summary of the finite temperature analysis of the FOPT, 
   and then present the resulting GW spectrum in the parameter regions that are complementary to Higgs precision measurements at the ILC.

% In the standard procedure for FOPT analysis, the thermally corrected effective scalar potential consists of 
%    a zero temperature effective potential with the one-loop CW potential \cite{Coleman:1973jx}, 
%    a finite temperature contribution \cite{Dolan:1973qd}, and 
%    an additional temperature dependent Daisy diagram contribution \cite{Carrington:1991hz}:
   
In the standard procedure for analyzing a FOPT, the thermally corrected effective scalar potential is composed of three parts: 
   the zero-temperature effective potential including the one-loop Coleman-Weinberg (CW) corrections \cite{Coleman:1973jx}, 
   the finite-temperature contribution \cite{Dolan:1973qd}, and the additional temperature-dependent Daisy (ring) 
   diagram contribution \cite{Carrington:1991hz}:   
%%%%%%%%%%%%%%%%%%%%%%%%%%%%%%%%%%%%%%%%%%%%%%%%%%%%%%%%%%
\begin{equation}
    V_{eff} \left( \phi, T \right) = V_{tree}\left( \phi \right) + V_{CW} \left( \phi \right) + V_{T}\left(\phi, T \right) + V_{daisy}\left(\phi, T \right), 
\end{equation}
%%%%%%%%%%%%%%%%%%%%%%%%%%%%%%%%%%%%%%%%%%%%%%%%%%%%%%%%%%
where 
%%%%%%%%%%%%%%%%%%%%%%%%%%%%%%%%%%%%%%%%%%%%%%%%%%%%%%%%%%
\begin{align}
       V_{CW} \left( \phi \right) &= \sum_{a} \eta_{a} n_{a} \frac{m_{a}^{4}\left( \phi \right)}{64 \pi^{2}} \Bigg[ log \frac{m_{a}^{2}\left( \phi \right)}{\Lambda^{2}} - C_{a}\Bigg],\\
       V_{T}\left(\phi, T \right) &= \frac{T^{4}}{2 \pi^2} \sum_{a} \eta_{a} n_{a} J_{b/f} \left( \frac{m_{a}^{2} \phi}{T^{2}} \right) ,\\
      % &\approx T^{2} \Bigg[ \sum_{b} \frac{n_{b}}{24} m_{b}^{2} \left( \phi \right) + \sum_{f} \frac{n_{f}}{48} m_{f}^{2} \left( \phi \right) \Bigg], \\
       V_{daisy}\left(\phi, T \right) &= -\frac{T}{12 \pi} \sum_{b} n_{b}^{L} \Bigg[ \left( m^{2}\left( \phi \right) + \Pi \left( T \right) \right)_{b}^{3/2} - \left( m^{2}\left( \phi \right)_{b}^{3/2} \right) \Bigg].
\end{align}
%%%%%%%%%%%%%%%%%%%%%%%%%%%%%%%%%%%%%%%%%%%%%%%%%%%%%%%%%%
Here, $V_{CW} \left( \phi \right)$ is written in a common alternative form, and
%%%%%%%%
\begin{align*}
    n_{a} &= \quad \text{\# DOF},\\
    n_{b/f} &= \quad \text{\# bosonic/fermionic DOF},\\
    n_{b}^{L} &= \quad \text{\# longitudinal bosonic DOF},\\
    \eta_{a} &= 
    \begin{cases}
        +1 & \text{(scalars)} \\
        -1 & \text{(fermions)}
    \end{cases}, \\
    m_{a}^{2}(\phi) &= \text{Field-dependent masses of scalars/fermions/bosons},\\
    \Lambda &= \quad \text{Renormalization scale},\\
    C_{a} &= 
    \begin{cases}
        \tfrac{3}{2} & \text{(scalars \& fermions)} \\
        \tfrac{5}{6} & \text{(gauge bosons)}
    \end{cases},\\
    J_{b/f}(x^{2}) &= \int_{0}^{\infty} dy \, y^{2} \, 
        \log \left( 1 \mp \exp\left[-\sqrt{y^{2} + x^{2}} \right] \right),
\end{align*}
%%%%%%%%%%%%%
where $J_{b/f} \left( x^{2} \right)$ are well-known bosonic/fermionic thermal functions.

Contributions from the bosonic function $J_{b}\left(x^{2}\right)$ in $V_{T}(\phi, T)$ generate a potential barrier 
  between the symmetric (high-temperature) and broken (low-temperature) vacua as $T \rightarrow 0$, 
  providing the conditions for a strong FOPT that can produce GWs. 
At the same time, contributions from $V_{daisy}$ tend to reduce the barrier, 
  since they carry the opposite sign relative to $V_{T}$. 
However, this cancellation is incomplete for the transverse gauge degrees of freedom, which do not contribute to $V_{daisy}$. 
Consequently, models with gauge-charged scalar fields, such as the conformal $U(1)_H$ Higgs sector considered here, are strong candidates for exhibiting a FOPT in the early universe, with a signature potentially observable at future GW experiments.
In a more general or extensive treatment, one must account for the gauge-dependence of $V_{eff} ( \phi, T )$ \cite{Lofgren:2021ogg, Liu:2026ask, Feng:2026iqo}.
The contributions to $V_{eff} ( \phi, T )$ calculated and used in this work are performed in the Landau gauge, for which Fadeev-Popov ghost terms do not contribute at one-loop level, so we do not consider the effects of gauge-dependence further here.

The FOPT proceeds analogously to a conventional thermodynamic phase transition, for example, a superheated liquid transitioning 
  to the vapor phase via nucleating bubbles. 
In the cosmological context, bubbles with nonzero VEVs nucleate at random locations in the universe and expand, 
   eventually colliding and filling space with the broken phase. 
As these bubbles expand and collide, they generate GWs through bubble collisions, turbulence, 
   and post-collision sound waves \cite{Caprini:2015zlo}. 
We now consider the parameters determined by $V_{eff}$ and their influence on these contributions to the GW signal.

%%%%%%%%%%%%%%%%%%%%%%%%%%%%%%%%%%%%%%%%%%%%%%%%%%%%%%%%%%
\subsection{GW parameters influenced by the scalar potential}
%%%%%%%%%%%%%%%%%%%%%%%%%%%%%%%%%%%%%%%%%%%%%%%%%%%%%%%%%%

The bubble \textit{nucleation rate} at finite temperature is given by 
%%%%%%%%%%%%%%%%%%%%%%%%%%%%%%%%%%%%%%%%%%%%%%%%%%%%%%%%%%
\begin{equation}
    \Gamma \left( T \right) = \Gamma_{0} e^{- S[\phi,T]} \simeq \Gamma_{0} e^{- S_{3}[\phi,T]/T} , 
\end{equation}
%%%%%%%%%%%%%%%%%%%%%%%%%%%%%%%%%%%%%%%%%%%%%%%%%%%%%%%%%%
where $\Gamma_{0}$ is a coefficient of order of the transition scale, $S$ is the action in the 4-dimensional Minkowski space, and $S_{3}$ is the 3-dimensional Euclidean action:
%%%%%%%%%%%%%%%%%%%%%%%%%%%%%%%%%%%%%%%%%%%%%%%%%%%%%%%%%%
\begin{equation}
    S[\phi,T] = \frac{S_{3}[\phi,T]}{T} = \frac{1}{T} \int \mathrm{d^{3}x} \Bigg[ \frac{1}{2} \left( \nabla \phi \right)^{2} + V_{eff} \left( \phi, T \right) \Bigg].
\end{equation}
%%%%%%%%%%%%%%%%%%%%%%%%%%%%%%%%%%%%%%%%%%%%%%%%%%%%%%%%%%
Demanding stationary action yields the O(3)-symmetric bounce equation, 
%%%%%%%%%%%%%%%%%%%%%%%%%%%%%%%%%%%%%%%%%%%%%%%%%%%%%%%%%%
\begin{equation}
    \frac{d^{2}\phi}{dr^{2}} + \frac{2}{r} \frac{d\phi}{dr} = V_{eff}^\prime \left( \phi, T \right)   ,
\end{equation}
%%%%%%%%%%%%%%%%%%%%%%%%%%%%%%%%%%%%%%%%%%%%%%%%%%%%%%%%%%
with boundary conditions $\phi \left( r\rightarrow \infty \right) \rightarrow 0$ and $\phi'\left( r=0 \right)=0$.
Solving this equation, we determine the nucleation rate.

The \textit{nucleation temperature} $T_{n}$ marks the start of the FOPT as the point where nucleating bubbles successfully expand instead of collapsing.
The value is obtained by requiring that the integrated number of bubbles generated in a Hubble volume $H^{-3} \sim 1$, or when
%%%%%%%%%%%%%%%%%%%%%%%%%%%%%%%%%%%%%%%%%%%%%%%%%%%%%%%%%%
\begin{equation} \label{eq:codition}
    \frac{S_{3}}{T} \sim 140 - 4 \, \textrm{log} \Bigg[ \frac{T_{n}}{100 \, \mathrm{GeV}} \Bigg]
\end{equation}
%%%%%%%%%%%%%%%%%%%%%%%%%%%%%%%%%%%%%%%%%%%%%%%%%%%%%%%%%%
is satisfied.

The \textit{transition strength} $\alpha$ of a FOPT is determined by the amount of released energy during the transition normalized with respect to the total energy density of the unbroken phase at the time of collision: 
%%%%%%%%%%%%%%%%%%%%%%%%%%%%%%%%%%%%%%%%%%%%%%%%%%%%%%%%%%
\begin{equation}
    \alpha \equiv \frac{\epsilon}{\rho_{rad}} = \frac{1}{\rho_{rad}} \left( -\Delta V_{\mathrm{eff}} 
    + T_{n} \left. \frac{\partial V_{\mathrm{eff}}}{\partial T} \right) \right|_{T_{n}} ,
\end{equation}
%%%%%%%%%%%%%%%%%%%%%%%%%%%%%%%%%%%%%%%%%%%%%%%%%%%%%%%%%%
where $\rho_{rad} = \frac{\pi^{2}}{30}  g_{*} T^{4}$.

Lastly, the \textit{transition timescale} for the FOPT is defined as the dimensionless parameter: 
%%%%%%%%%%%%%%%%%%%%%%%%%%%%%%%%%%%%%%%%%%%%%%%%%%%%%%%%%%
\begin{equation}
    \frac{\beta}{H} \equiv \left.  T_{n} \frac{\mathrm{d}}{\mathrm{dT}} \left( \frac{S_{3}}{T} \right) \right|_{T_{n}} ,
\end{equation}
%%%%%%%%%%%%%%%%%%%%%%%%%%%%%%%%%%%%%%%%%%%%%%%%%%%%%%%%%%
which depends on the shape of $V_{eff}$ at the time of nucleation.

%%%%%%%%%%%%%%%%%%%%%%%%%%%%%%%%%%%%%%%%%%%%%%%%%%%%%%%%%%
\subsection{The GW spectrum}
%%%%%%%%%%%%%%%%%%%%%%%%%%%%%%%%%%%%%%%%%%%%%%%%%%%%%%%%%%

The GW Spectrum itself consists of the three previously mentioned contributions from bubble collisions, turbulence, 
   and sound waves \cite{Caprini:2015zlo}:
%%%%%%%%%%%%%%%%%%%%%%%%%%%%%%%%%%%%%%%%%%%%%%%%%%%%%%%%%%
\begin{equation}
    \Omega_{GW} \left( f \right) = \Omega_{\phi} \left( f \right) + \Omega_{sw} \left( f \right) + \Omega_{turb} \left( f \right)  .
\end{equation}
%%%%%%%%%%%%%%%%%%%%%%%%%%%%%%%%%%%%%%%%%%%%%%%%%%%%%%%%%%

%%%%%%%%%%%%%%%%%%%%%%%%%%%%%%%%%%%%%%%%%%%%%%%%%%%%%%%%%%
\subsubsection{$\Omega_{\phi}$ - Bubble Collisions}
For GW contributions from bubble collisions, the peak frequency and peak amplitude of the signal are given by
%%%%%%%%%%%%%%%%%%%%%%%%%%%%%%%%%%%%%%%%%%%%%%%%%%%%%%%%%%
\begin{align}
    f_{peak} &\simeq 17 \left( \frac{f_{*}}{\beta} \right) \left( \frac{\beta}{H_{*}} \right) \left( \frac{T_{*}}{10^{8} \textrm{GeV}} \right) \left( \frac{g_{*}}{100} \right)^{1/6} \textrm{Hz} , \\
    h^{2} \Omega_{\phi} \left( f_{peak} \right) &= 1.7 \times 10^{-5} \kappa^{2} \Delta \left( \frac{\beta}{H_{*}} \right)^{-2} \left( \frac{\alpha}{1 + \alpha} \right)^{2} \left( \frac{g_{*}}{100} \right)^{-1/3} ,
\end{align}
%%%%%%%%%%%%%%%%%%%%%%%%%%%%%%%%%%%%%%%%%%%%%%%%%%%%%%%%%%
where 
%%%%%%%%%%%%%%%%%%%%%%%%%%%%%%%%%%%%%%%%%%%%%%%%%%%%%%%%%%
\begin{align}
    \Delta &= \frac{0.11 v_{b}^{3}}{0.42 + v_{b}^{2}} , \\ 
    \frac{f_{*}}{\beta} &= \frac{0.62}{1.8 - 0.1 v_{b} + v_{b}^{2}} ,
\end{align}
%%%%%%%%%%%%%%%%%%%%%%%%%%%%%%%%%%%%%%%%%%%%%%%%%%%%%%%%%%
with bubble wall velocity $v_{b}$. 
The efficiency factor $\kappa$ is given by 
%%%%%%%%%%%%%%%%%%%%%%%%%%%%%%%%%%%%%%%%%%%%%%%%%%%%%%%%%%
\begin{equation}
    \kappa = \frac{1}{1 + A \alpha} \left( A \alpha + \frac{4}{27} \sqrt{\frac{3\alpha}{2}} \right)
\end{equation}
%%%%%%%%%%%%%%%%%%%%%%%%%%%%%%%%%%%%%%%%%%%%%%%%%%%%%%%%%%
with $A = 0.715$.
The overall spectrum is well-approximated by
%%%%%%%%%%%%%%%%%%%%%%%%%%%%%%%%%%%%%%%%%%%%%%%%%%%%%%%%%%
\begin{equation}
    \Omega_{\phi} \left( f \right) = \Omega_{\phi} \left( f_{peak} \right) \frac{\left( a + b \right) f_{peak}^{b} f^{a}}{b f_{peak}^{a+b} + a f^{a + b}}.
\end{equation}
%%%%%%%%%%%%%%%%%%%%%%%%%%%%%%%%%%%%%%%%%%%%%%%%%%%%%%%%%%
In our analysis, we use $\left( a, b, v_{b} \right) = \left( 2.7, 1.0, 1.0 \right)$.

\subsubsection{$\Omega_{sw}$ - Sound waves}
For sound wave contributions, the peak frequency and amplitude are given by
%%%%%%%%%%%%%%%%%%%%%%%%%%%%%%%%%%%%%%%%%%%%%%%%%%%%%%%%%%
\begin{align}
    f_{peak} &\simeq 19 \frac{1}{v_{b}} \left( \frac{\beta}{H_{*}} \right) \left( \frac{T_{*}}{10^{8} \textrm{GeV}} \right)
     \left( \frac{g_{*}}{100} \right)^{1/6} \textrm{Hz}  , \\
    h^{2} \Omega_{SW} \left( f_{peak} \right) &\simeq 2.7 \times 10^{-6} \kappa_{v}^{2} v_{b} 
    \Delta \left( \frac{\beta}{H_{*}} \right)^{-1} \left( \frac{\alpha}{1 + \alpha} \right)^{2} \left( \frac{g_{*}}{100} \right)^{-1/3}.
\end{align}
%%%%%%%%%%%%%%%%%%%%%%%%%%%%%%%%%%%%%%%%%%%%%%%%%%%%%%%%%%
The efficiency factor $\kappa_{v}$ is given by
%%%%%%%%%%%%%%%%%%%%%%%%%%%%%%%%%%%%%%%%%%%%%%%%%%%%%%%%%%
\begin{equation}
    \kappa_{v} \simeq \frac{\alpha}{0.73 + 0.083 \sqrt{\alpha} + \alpha}  \quad \text{for} \quad   v_{b} \simeq 1.
\end{equation}
%%%%%%%%%%%%%%%%%%%%%%%%%%%%%%%%%%%%%%%%%%%%%%%%%%%%%%%%%%
The total spectrum contribution is fitted by 
%%%%%%%%%%%%%%%%%%%%%%%%%%%%%%%%%%%%%%%%%%%%%%%%%%%%%%%%%%
\begin{equation}
    \Omega_{SW} \left( f \right) = \Omega_{SW} \left( f_{peak} \right) \left( \frac{f}{f_{peak}} \right)^{3} \left( \frac{7}{4 + 3 \left( \frac{f}{f_{peak}} \right)^{2}} \right)^{7/2}.
\end{equation}
%%%%%%%%%%%%%%%%%%%%%%%%%%%%%%%%%%%%%%%%%%%%%%%%%%%%%%%%%%

\subsubsection{$\Omega_{turb}$ - Turbulence }
For contributions from turbulence, the peak frequency and amplitude are given by
%%%%%%%%%%%%%%%%%%%%%%%%%%%%%%%%%%%%%%%%%%%%%%%%%%%%%%%%%%
\begin{align}
    f_{peak} &\simeq 27 \frac{1}{v_{b}} \left( \frac{\beta}{H_{*}} \right) 
    \left( \frac{T_{*}}{10^{8} \textrm{GeV}} \right) \left( \frac{g_{*}}{100} \right)^{1/6} \textrm{Hz} , \\
    h^{2} \Omega_{turb} \left( f_{peak} \right) &\simeq 3.4 \times 10^{-4} v_{b} \left( \frac{\beta}{H_{*}} \right)^{-1} \left( \frac{\kappa_{turb} \alpha}{1 + \alpha} \right)^{3/2} \left( \frac{g_{*}}{100} \right)^{-1/3}.
\end{align}
%%%%%%%%%%%%%%%%%%%%%%%%%%%%%%%%%%%%%%%%%%%%%%%%%%%%%%%%%%
In accordance with Ref.~\cite{Caprini:2015zlo}, we take $\kappa_{turb} \simeq 0.05 \kappa_{v}$.
The spectrum shape is then fitted by
%%%%%%%%%%%%%%%%%%%%%%%%%%%%%%%%%%%%%%%%%%%%%%%%%%%%%%%%%%
\begin{equation}
    \Omega_{turb} \left( f \right) = \Omega_{turb} \left( f_{peak} \right) \frac{\left( \frac{f}{f_{peak}}  \right)^{3}}{\left( 1 + \frac{f}{f_{peak}} \right)^{11/3} \left( 1 + \frac{8 \pi f}{h_{*}} \right)}
\end{equation}
%%%%%%%%%%%%%%%%%%%%%%%%%%%%%%%%%%%%%%%%%%%%%%%%%%%%%%%%%%
with
%%%%%%%%%%%%%%%%%%%%%%%%%%%%%%%%%%%%%%%%%%%%%%%%%%%%%%%%%%
\begin{equation}
    h_{*} = 17 \left( \frac{T_{*}}{10^{8} \, \text{GeV}} \right) \left( \frac{g_{*}}{100} \right)^{1/6} \textrm{Hz}.
\end{equation}
%%%%%%%%%%%%%%%%%%%%%%%%%%%%%%%%%%%%%%%%%%%%%%%%%%%%%%%%%%

%%%%%%%%%%%%  Fig  %%%%%%%%%%%%%%%%%%%%%%%%%%%%%%%%%%%%%%
\begin{figure}[!ht]
\centering
%[width = 0.47\textwidth]
\includegraphics[scale=0.9]{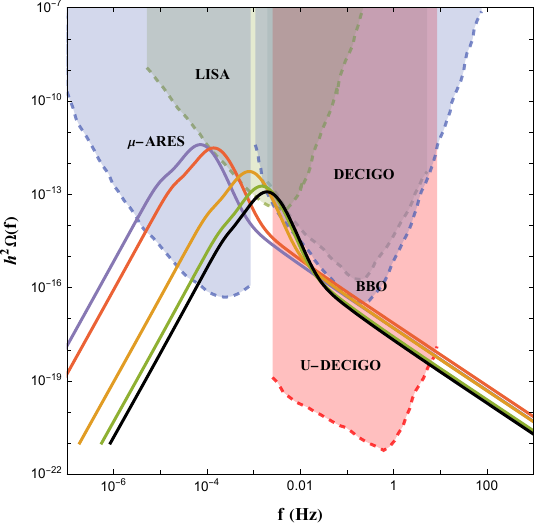}
\caption{
Gravitational wave spectra for $M_{h_{2}} =$30 GeV (purple), 35 GeV (red), 40 GeV (orange), 45 GeV (green), and 50 GeV (black) 
  in our conformal model, evaluated at the largest $g_{H}$ values allowed by LEP-II bounds. 
While the peak signals are only marginally accessible to near-future interferometers, 
  they fall well within the projected sensitivities of far-future detectors such as Ultimate-DECIGO \cite{Kudoh:2005as} and $\mu$-ARES \cite{Sesana:2019vho}.
}
\label{fig:GWparamspace}
\end{figure}
%%%%%%%%%%%%%%%%%%%%%%%%%%%%%%%%%%%%%%%%%%%%%%%%%%%%%

%%%%%%%%%%%%%%%%%%%%%%%%%%%%%%%%%%%%%%%%%%%%%%%%%%%%%%%%%%
\subsection{GWs from the Classically Conformal $U(1)_H$ extended SM}
%%%%%%%%%%%%%%%%%%%%%%%%%%%%%%%%%%%%%%%%%%%%%%%%%%%%%%%%%%

Following the prescriptions for finite-temperature potential analysis and GW signal estimation,
   we evaluate the GW spectra for our conformal model.
Our study does not involve the full two-dimensional Higgs potential, but focuses on the $\phi$-direction only. 
This simplification is justified by the strong hierarchy between $v_{h} = 246$ GeV and $v_{\phi} = 2$-$6$ TeV 
  (corresponding to the lines shown in Fig.~\ref{fig:BrOfSinsq}), 
  together with the very small mixed quartic coupling. 
Furthermore, we neglect the contribution of the $\phi$ self-coupling ($\lambda_\phi$) to $V_{daisy}$, since $\lambda_\phi \ll g_H$.

Considering the stringent constraints from direct DM detection, we abandon the $Z^\prime$ DM scenario 
   and instead focus on a set of parameter choices, shown in the left panel of Fig.\ref{fig:Zpparamspace} without the purple line, 
   that yield a FOPT and an associated GW signal.
In general, the nucleation condition in Eq.~(\ref{eq:codition}) is more easily satisfied when the gauge coupling $g_H$ is larger, 
   though there is also a correlated dependence on $v_{\phi}$, which varies with $g_{H}$.
In the specific case of our conformal model, we find that gauge couplings near $g_{H} \simeq 0.2$, with corresponding 
   $v_{\phi} = \mathcal{O}$(1 TeV), most reliably satisfy the nucleation conditions and produce successful FOPTs. 
We consider five representative choices of $M_{h_{2}}$ = 30, 35, 40, 45, and 50 GeV, and compute the GW spectrum 
   using the LEP-II upper bound on $g_H$ for each case.

Figure~\ref{fig:GWparamspace} shows the resulting GW spectra together with the prospective sensitivity regions 
   of future GW detection experiments.
For all five cases, we obtain peak amplitudes in the range $h^{2} \Omega_{GW}(f_{peak}) = 10^{-13}$-$10^{-12}$,
    at peak frequencies of $f_{peak} = 10^{-5}$-$10^{-3}$ Hz. 
As shown in Fig.~\ref{fig:GWparamspace}, the predicted spectra are mostly outside the reach of near-future detectors 
   such as LISA \cite{LISA:2017pwj}, DECIGO\cite{Kawamura:2006up}, and BBO\cite{Harry:2006fi}, but they fall within the sensitivity ranges of more advanced detectors such as Ultimate-DECIGO \cite{Kudoh:2005as}
   and the recently proposed $\mu$-ARES \cite{Sesana:2019vho}.
These GW signals are complementary to precision Higgs measurements at the ILC.

%%%%%%%%%%%%%%%%%%%%%%%%%%%%%%%%%%%%%%%%%%%%%%%%%%%%%%%%
\section{Conclusion}
\label{sec:5}
%%%%%%%%%%%%%%%%%%%%%%%%%%%%%%%%%%%%%%%%%%%%%%%%%%%%%%%%

We have studied a classically conformal $U(1)_H$ extension of the Standard Model (SM), 
   in which the $U(1)_{H}$ Higgs field $\Phi$ with charge $+2$ triggers radiative symmetry breaking via the Coleman-Weinberg mechanism. 
This generates a negative mass squared for the SM Higgs doublet through the mixed quartic coupling, 
  thereby inducing electroweak (EW) symmetry breaking. 
In this framework, the origin of EW symmetry breaking is directly linked to the radiative breaking of the hidden $U(1)_H$ gauge symmetry.

A key phenomenological feature of this scenario is the strong suppression of the $h_1 \to h_2 h_2$ decay 
   in the case of $M_{h_1} > 2 M_{h_2}$, where $h_1$ and $h_2$ are the SM-like and SM singlet-like Higgs bosons.  
Consequently, the anomalous Higgs decay $h_1 \to h_2 h_2$ followed by $h_2\to b \bar{b}$ expected in conventional Higgs portal models 
   is highly suppressed here. 
This implies that, unlike in the conventional case where both anomalous couplings and anomalous decays can be probed simultaneously at the ILC, 
   in the conformal case only the anomalous couplings remain within reach while the decay channel effectively evades detection. 
This characteristic provides an important discriminator for the conformal origin of EW symmetry breaking, 
  to be tested at future lepton colliders such as the ILC or MuC. 

As an illustration of experimental complementarity, we examined two scenarios beyond collider probes. 
First, we considered the $U(1)_H$ gauge boson as the sole dark matter (DM) candidate in the Universe.  
While the model can reproduce the observed relic abundance along a specific $g_H$-$m_{Z'}$ relation 
   that overlaps with the Higgs precision frontier, the relevant parameter space is excluded by the recent LZ results. 
Second, we investigated the possibility of a strong first-order phase transition (FOPT) associated with $U(1)_H$ breaking in the early universe. 
For parameter choices relevant to future Higgs precision tests, we found that the resulting stochastic gravitational wave background 
   lies largely beyond the reach of near-future observatories such as LISA and BBO, 
   but can be probed by far-future facilities including $\mu$-ARES and Ultimate DECIGO.

In summary, the classically conformal $U(1)H$ extension of the SM with $M{_{h_1}} > 2 M_{h_2}$ 
  exhibits distinctive collider signatures - suppressed anomalous Higgs decays alongside observable anomalous couplings - 
  as well as complementary gravitational wave signals. 
Together, future Higgs precision measurements and gravitational wave searches provide a powerful avenue 
  to test the conformal origin of electroweak symmetry breaking.

%%%%%%%%%%%%%%%%%%%%%%%%%%%%%%%%%%
\section*{Acknowledgments}
%%%%%%%%%%%%%%%%%%%%%%%%%%%%%%%%%%
This work is supported in part by the United States Department of Energy Grants DE-SC-0012447 (N.O.) 
and DE-SC0023713 (B.V. and N.O.).

%%%%%%%%%%%%%%%%%%%%%%%%%%%  
\bibliographystyle{utphysII}
\bibliography{References}

\end{document}